\documentclass[aps,prl, twocolumn, showpacs,amsmath prepintnumbers,amssymb,superscriptaddress,reprint]{revtex4}

\usepackage[dvipdfmx]{graphicx}
\usepackage{dcolumn}
\usepackage{color}
\usepackage[latin1]{inputenc}
\usepackage{comment}
\usepackage{amsmath}
\usepackage{here}
\usepackage[normalem]{ulem}

\begin{document}

\title
  {Ultrafast Energy- and Momentum-resolved Surface Dirac Photocurrents in the Topological Insulator Sb$_{2}$Te$_{3}$}

\author{Kenta Kuroda}
\affiliation{Institute for Solid State Physics (ISSP), University of Tokyo, 277-8581 Chiba, Japan}
\affiliation{Fachbereich
Physik und Zentrum f{\"u}r Materialwissenschaften,
Philipps-Universit{\"a}t, 35032 Marburg, Germany}
\email{kuroken224@issp.u-tokyo.ac.jp}
\author{J. Reimann}
\affiliation{Fachbereich
Physik und Zentrum f{\"u}r Materialwissenschaften,
Philipps-Universit{\"a}t, 35032 Marburg, Germany}
\author{K.~A.~Kokh}
\affiliation{Institute of Geology and Mineralogy, Siberian Branch, Russian Academy of Sciences, Koptyuga pr. 3, 630090 Novosibirsk, Russia}
\affiliation{Novosibirsk State University, ul. Pirogova 2, 630090 Novosibirsk, Russia}
\affiliation{Saint Petersburg State University, 198504 Saint Petersburg, Russia}
\author{O.~E.~Tereshchenko}
\affiliation{Novosibirsk State University, ul. Pirogova 2, 630090 Novosibirsk, Russia}
\affiliation{Saint Petersburg State University, 198504 Saint Petersburg, Russia}
\affiliation{Institute of Semiconductor Physics, Siberian Branch, Russian Academy of Sciences, pr. Akademika Lavrent'eva 13, 630090 Novosibirsk, Russia}
\author{A.~Kimura}
\affiliation{Graduate School of Science, Hiroshima University, 739-8526 Hiroshima, Japan}
\author{J.~G{\"u}dde}
\author{U.~H{\"o}fer}
\affiliation{Fachbereich
Physik und Zentrum f{\"u}r Materialwissenschaften,
Philipps-Universit{\"a}t, 35032 Marburg, Germany}

\date{\today}

\begin{abstract}
We present energy-momentum mapping of surface Dirac photocurrent in the topological insulator Sb$_{2}$Te$_{3}$ by means of time- and angle-resolved two-photon photoemission spectroscopy combined with polarization-variable mid-infrared pulse laser.
It is demonstrated that the direct optical transition from the occupied to the unoccupied part of the surface Dirac-cone permits the linear and circular photogalvanic effect which thereby enables us to coherently control the surface electric-current by laser polarization.
Moreover, the photocurrent mapping directly visualizes ultrafast current dynamics in the Dirac cone as a function of time.
We unravel the ultrafast intraband relaxation dynamics of the inelastic scattering and momentum scattering separately.
Our observations pave the pathway for coherent optical control over surface Dirac electrons in topological insulators.
\end{abstract}
\maketitle
%
%
%
A realization of functional capabilities to generate and control highly spin-polarized electrons is a central subject in the research for spintronic application~\cite{Manchon15NatureMat}.
As a promising spintronic material, spin-polarized topological surface state (TSS) in three-dimensional topological insulators (TIs) has grabbed particular attentions~\cite{,Hasan10rmp,Ando13jpsj}.
The TSS forms Dirac-cone-like energy dispersion~\cite{Xia09Nature,Chen09Science} and exhibits chiral spin texture in momentum space due to the strong spin-orbit coupling (SOC)~\cite{Heish09Nature,Pan_PRL,Souma_spin,Miyamoto_spin}.
In recent years, the TIs face a new challenge for an optical control over a net spin-polarized electric current of the TSS in out-of-equilibrium.
It was shown that a polarization-dependent asymmetric optical excitation of the TSSs in momentum-space, the so-called photogalvanic effect, generates the photocurrent~\cite{McIver12NatureNano,Olbrich14PRL,Kastl15NatureCom,Ogawa16NatureCom}, suggesting optospintronic devices based on the TIs.
All these performances rely on the ultrafast excitation and dynamics of hot TSS carriers, which has been intensively investigated in a number of different experiments, optical reflectivity~\cite{Qi2010APL,Hsieh2011PRL}, optically triggered detection of photocurrent~\cite{Kastl15NatureCom}.
However, in these bulk sensitive experiments, the finite bulk responses generally govern the total signal and therefore pure TSS dynamics is hardly detected.
\begin{figure*}[t]
\begin{center}
\includegraphics[width=0.95\textwidth]{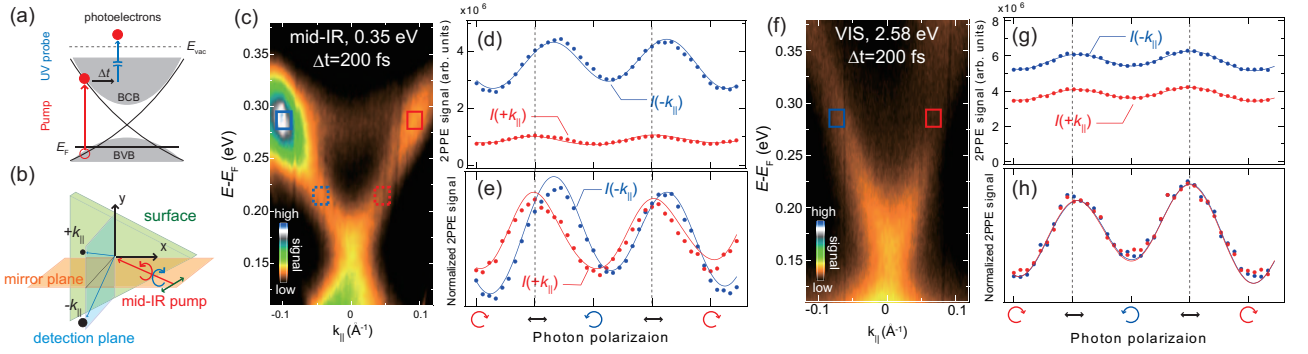}
\caption[]{(Color online)
(a) Excitation scheme for the transition into the TSS in 2PPE experiment.
(b) Experimental geometry where a mirror plane of the surface coincides with the plane of incidence (red square) perpendicular to the detection plane of photoelectrons (blue square).
The triangle depicts the $C_{3v}$ surface symmetry.
(c) and (f) Angle-resolved 2PPE spectra at 200~fs after temporal overlapping pump and probe pulses of Sb$_{2}$Te$_{3}$ by using $h\nu_{1}$=0.35~eV and 2.58 eV, respectively.
(d) and (g) Transient 2PPE intensities for different pump polarization within the integration windows around $E-E_{\rm{F}}$=0.28~eV denoted by solid line in (c) and (f).
(e) and (f) The corresponding normalized transient intensities, which are defined as $I$($\pm{k}_{||}$)/$I_{\rm{total}}$($\pm{k}_{||}$) where $I_{\rm{total}}$($\pm{k}_{||}$) is the integration intensity for a period of the mid-IR polarizations.
The data are shown with the fitting results with Equations~(\ref{eq1}) and (\ref{eq2}).
The phase shift can be see in (d) and (e), which is found to be $\sim$20$^\circ$ of $\lambda$/4-waveplate angle
}
\label{fig1}
\end{center}
\end{figure*}

Recently we have shown great capabilities of time- and angle-resolved two-photon photo emission spectroscopy (2PPE) combined with mid-infrared (mid-IR) pump pulses to generate and detect the surface photocurrent in a TI, Sb$_2$Te$_3$~\cite{Kuroda16prl}.
2PPE can visualize the photocurrent excitation and dynamics through the direct mapping of the asymmetric momentum-distribution of the optically excited electrons as a function of time~\cite{Guedde07Science}.
Moreover, in contrast to other 2PPE experiments on TSSs with higher excitation energy typically 1.55~eV~\cite{Sobota12prl,Wang12prl,Hajlaoui12Nano,Crepaldi12prb,Zhu15scire} and visible range~\cite{Niesner14prb,Reimann14prb}, the low energy excitation of mid-IR enables us to investigate the surface dynamics owing to a suppression of indirect excitations of the TSS from the higher-lying bulk states.
Surprisingly, it was found that the mid-IR excitation causes a direct optical excitation from the occupied to the unoccupied TSS across the Dirac point and the coherent optical transition moreover generates the surface photocurrent even by using linearly polarized light.
The next stage hence becomes a challenge for an optical control over the coherent optical transition and the resulting surface photocurrent.
Although this is generally believed to be achieved by using circularly polarized light~\cite{McIver12NatureNano,Kastl15NatureCom, Hosur11prb}, some THz experiments showed that the circular polarization plays only a minor role in generating the photocurrent~\cite{Plank16prb,Onishi14arxiv}.

In this Letter, by presenting an investigation of light polarization dependence in the coherent optical transition, we provide a direct evidence that the surface Dirac current can be coherently controlled through the linear and circular photogalvanic effect while the current direction is insensitive to the light polarization.
By monitoring the decay of the photocurrent in energy-momentum space, we furthermore uncover different intraband dynamics of the current in the surface Dirac-cone.

Details of our optical setup are described in the Supplemental Material~\cite{si}.
In our 2PPE measurements, electrons were excited into initially unoccupied states with mid-IR laser pulses ($h\nu_{1}$=0.35~eV, 100~fs) or visible laser pulses ($h\nu_{1}$=2.58~eV, 80~fs).
The polarization of these pump pulses was controlled by $\lambda$/4-waveplate, from horizontal linear polarization (HR) to right- or left-circular polarization.
The polarization-dependent populations in the transients was subsequently probed by photoemission of these electrons using ultraviolet (UV) laser pulses ($h\nu_{2}$=5.16~eV, 80~fs) [Fig.~\ref{fig1}(a)].
The experiments were carried out in a $\mu$-metal shielded UHV chamber at a base pressure of 4$\times$10$^{-11}$~mbar.
Photoelectrons were collected by a hemispherical analyzer (Specs Phoibos 150) with a display-type detector and along the high symmetry line $\bar\Gamma$-$\bar{K}$ [see Fig.~\ref{fig1}(b)].
Single crystal of $p$-doped Sb$_2$Te$_3$ were cleaved in situ by the Scotch tape method at a pressure of 3$\times$10$^{-10}$~mbar followed by a rapid recovery back to the base pressure within a minute.
The Dirac point (DP) in the sample is located around 150 meV above the Fermi energy ($E_{\rm{F}}$).
During the measurements the sample temperature was maintained at 80 K.

Figure~\ref{fig1}(c) displays a representative angle-resolved 2PPE intensity maps at 200~fs after temporal overlap-ping by using $h\nu_{1}$=0.35~eV for HR.
The population of the TSS is pronounced at a specific energy around 280~meV due to the direct optical transition from the lower part of the occupied TSS, and it shows much more prominent for $-k_{||}$ compared to that for $+k_{||}$ reflecting the transient TSS current excited by using HR~\cite{Kuroda16prl}.
As a heart of our interests, we monitor the direct population by changing polarization of mid-IR pump pulses.
Figure~\ref{fig1}(d) represents the polarization evolution of the direct population for $\pm{k}_{||}$, which are integrated within the windows denoted by solid line in Fig.
The data shows that the both populations oscillate as a function of the polarization and the population for $-k_{||}$ is always larger for any polarization.
The polarization of mid-IR, thus, plays only a minor role in the direction of the TSS photocurrent.
\begin{figure}[b]
\begin{center}
\includegraphics[width=0.9\columnwidth]{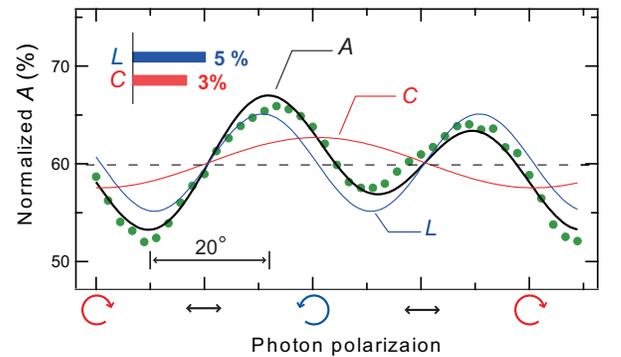}%
\caption[]{(Color online) Pump polarization dependence of the normalized differential intensity ($A$) that is defined by Equation~(\ref{eq1}).
The data points (denoted with closed circles) are obtained from the data of Figure~\ref{fig1}(e).
The red and blue lines are fitting results for the surface photocurrent due to circular and linear photogalvanics effects, respectively.
The angle difference in $\lambda$/4-waveplate angle between the photocurrent minimum and maximum is found to be nearly 50$^\circ$.
}
\label{fig2}
\end{center}
\end{figure}
\begin{figure*}[t]
\begin{center}
\includegraphics[width=0.9\textwidth]{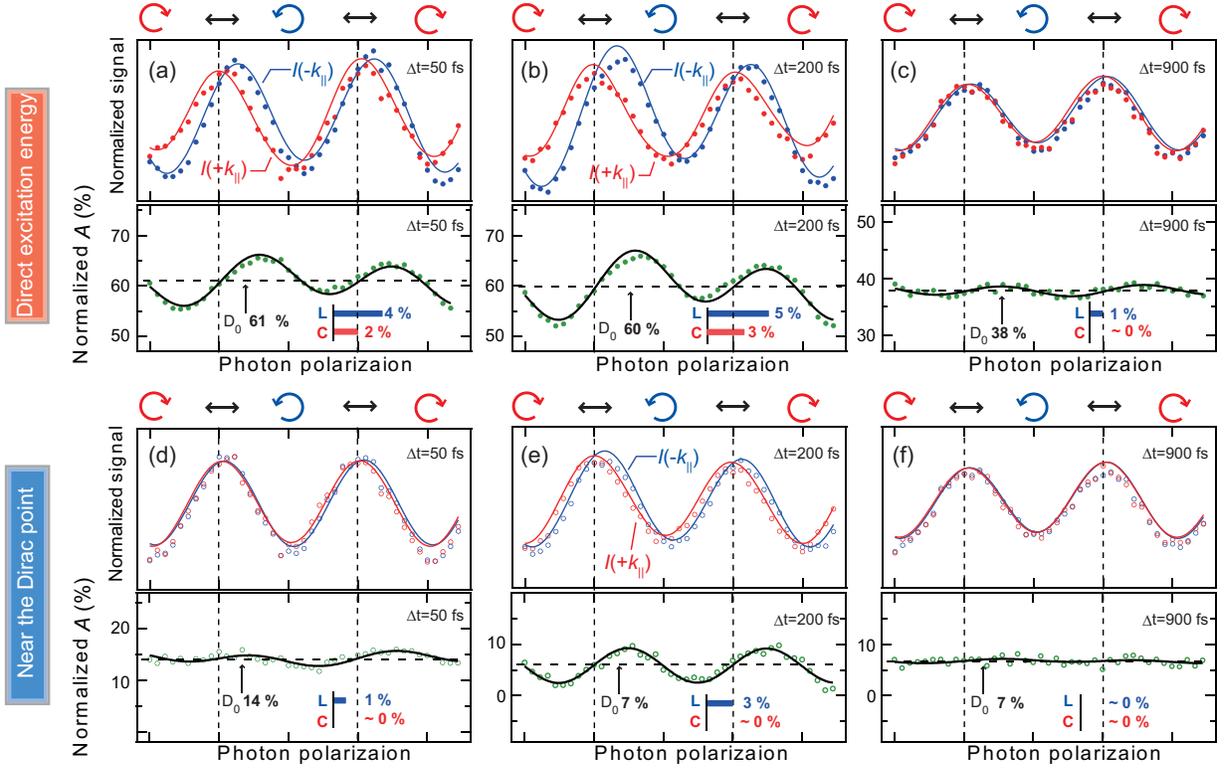}%
\caption[]{(color online) Time and energy evolution of the surface photocurrents.
The normalized transient intensities for the TSS at the direct excitation energy [(a)-(c)] and near the Dirac point [(d)-(f)], which are both integrated within the energy windows shown in figure 1(c).
The data are selectively shown at several representative delay-time of pump and probe pulses ($\Delta{t}$=50, 200, 900~fs).
The bottom figures represents the corresponding polarization dependence of the surface photocurrent.
The data are shown with the fitting results with Equation~(\ref{eq1}) and (\ref{eq2}).
}
\label{fig3}
\end{center}
\end{figure*}
To clarify the polarization dependence for $\pm{k}_{||}$, which is superimposed on the strongly asymmetric population, we show in Fig.~\ref{fig1}(e), the 2PPE signal in which the oscillation amplitude of Fig.~\ref{fig1}(d) are normalized by the total intensity [see the caption of Fig.~\ref{fig1}].
We find a clear difference in the polarization evolutions for $\pm{k}_{||}$ as a phase shift: the population maximum for $-k_{||}$ ($+k_{||}$) shifts to right (left) with respect to the angle of $\lambda$/4-waveplate for HR.
In sharp contrast to these results for mid-IR excitation, the strong asymmetric population and the phase shift are not seen in the 2.58~eV excitation as shown in Figs.~\ref{fig1}(f)-(h), in which the populations are mostly indirect fillings~\cite{Reimann14prb}.
Therefore, the both are unique products of the coherent optical transition of mid-IR pulse, indicating the presence of the polarization-driven photocurrent.

We now turn to describe a more detailed analysis of the polarization driving photocurrent with the population asymmetry ($A$) defined as
\begin{equation}
A=\frac{I_{-k||}-I_{+k||}}{I_{-k||}+I_{+k||}}.
\label{eq1}
\end{equation}
where $I_{\pm{k}_{||}}$ are the 2PPE signal for $\pm{k}_{||}$.
Figure~\ref{fig2} represents $A$ of the direct population.
We observe a small polarization dependence within $\pm$8~$\%$ and a polarization independent component of 60~$\%$ related to the strongly asymmetric population as clearly seen in Fig.~\ref{fig1}(d).
The polarization dependence of $A$ precisely indicates the presence of the phase shift observed in the transient electron populations in Fig.~\ref{fig1}(e). 
Remarkably, the value of $A$ becomes the maximum or minimum when the light polarization is tuned to be in between circular polarization and HR.
The relative angle of $\lambda$/4-waveplate for producing the maximum and minimum $A$ is found to be nearly 50$^{\circ}$.
This cannot be explained only by the helicity dependence that must induce 90~$^{\circ}$ of the phase shift.
In fact, the observed polarization dependence away from the simple helicity dependence was also reported in the previous photocurrent transport experiments suggesting superposition of linear and circular photogalvanic effect~\cite{McIver12NatureNano,Kastl15NatureCom}.

To decompose the linear and circular photogalvanic effect, we describe the surface photocurrent as follows:
\begin{equation}
A=Ccos2\theta + Lcos4\theta + D.
\label{eq2}
\end{equation}
The coefficient $C $and $L$ parameterize the helicity- and the linear-polarization-dependent photocurrent, respectively.
The third term ($D$) evaluates the polarization-independent photocurrent.
The result of the fitting analysis with Equation~(\ref{eq2}) is shown in Figure~\ref{fig2}.
Clearly, this simple model can reproduce the polarization-dependent $A$ and even the phase shift of the transient population as shown in Figs.~\ref{fig1}(d) and (e).
We find that the linear polarization of the mid-IR pump laser contributes to the surface photocurrent rather than the circular polarization.
We attribute the large linear polarization dependence to the elliptical polarization radiation~\cite{Plank16prb}.
These results directly demonstrate that the direct optical coupling of the TSS allows the photogalvanic effect and thereby enables us to coherently control the amplitude of the surface electric-current.

%
Figure~\ref{fig3} summarizes the polarization evolution of the normalized $I_{\pm{k}_{||}}$ and $A$ monitored at different energy and delay time ($\Delta{t}$).
Near the temporal overlapping ($\Delta{t}$=50~fs), we already see the clear phase shift and the resulting polarization dependence in the TSS photocurrent at the direct excitation energy ($E_{\rm{direct}}$) [Fig.~\ref{fig3}(a)], 
while these are negligibly small at energy around DP [Fig.~\ref{fig3}(d)] and in the bulk conduction bands~\cite{si}.
The results, therefore, indicate that the direct excitation is an important factor to coherently control the electric surface currents via the photogalvanic effect.
Interestingly, after 200~fs, the transient populations near the DP  start to display a clear phase shift [Fig.~\ref{fig3}(e)].
Finally, the phase shift at $E_{\rm{direct}}$ and near DP disappears after 900~fs although $D$ considerably remains 40~$\%$ at $E_{\rm{direct}}$ [Figs.~\ref{fig3}(c) and (f)].

\begin{figure}[t]
\begin{center}
\includegraphics[width=0.85\columnwidth]{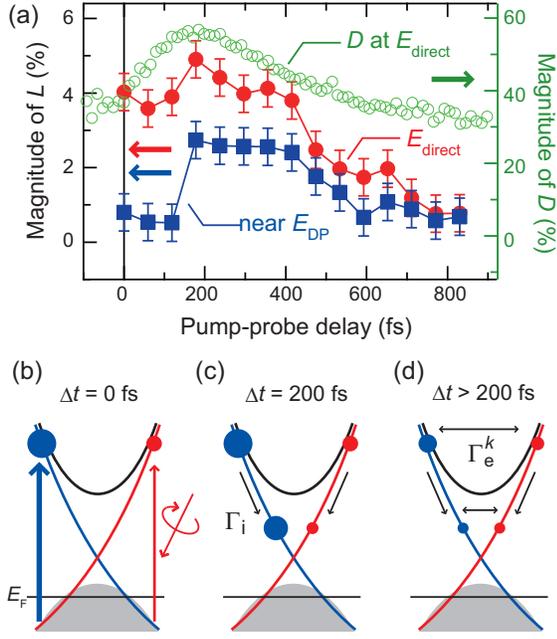}%
\caption[]{(Color online) (a) Temporal evolutions of the magnitude of linear polarization dependence $L$ (closed circles) at $E_{\rm{direct}}$ and (closed squares) near the Dirac point ($E_{\rm{DP}}$), which were both obtained from the fitting analysis from Fig.~\ref{fig3}.
These are compared to the polarization-independent photocurrent which is taken from our previously result~\cite{Kuroda16prl}.
(b)-(d) Schematic of the TSS photocurrent excitation by using mid-IR pump and its decay dynamics, showing (b) the direct optical transition of the TSS coherently generating the surface photocurrent, (c) the excited electron at $E_{\rm{direct}}$ populates the lower unoccupied Dirac-cone through inelastic intraband scattering ($\Gamma_{i}$) within 200~fs.
(d) The surface photocurrent decays through the intraband momentum scattering ($\Gamma_{e}^{k}$).
}
\label{fig4}
\end{center}
\end{figure}
In Fig.~\ref{fig4}(a), we plot the temporal evolution of $L$ at the two energy and these are compared to polarization-independent photocurrent at $E_{\rm{direct}}$.
The surface photocurrent excited at almost temporal overlap of pump and probe pulses at $E_{\rm{direct}}$.
Compared to the fast photocurrent excitation, it obviously delays near DP, and drastically increases around at 200~fs.
These results display the carrier dynamics of the TSS which is summarized in Figs.~\ref{fig4}(b)-(d).
The photocurrent is firstly excited due to the direct optical transition of the TSS at $\Delta{t}$=0~fs [Fig.~\ref{fig4}(b)]. Within 200~fs, the excited electrons starts to populate the unoccupied Dirac-cone at energies below generating the asymmetric electron population even near the DP as shown in Fig.~\ref{fig4}(c).
Importantly, the delayed fillings of $L$ indicates the time scale of inelastic intraband scattering in the surface Dirac-cone ($\Gamma_{i}$), which is usually hindered by strong indirect filling of the electrons excited in the bulk conduction bands~\cite{Sobota12prl}.
After 200~fs, the photocurrent decays through the elastic momentum scattering ($\Gamma_{e}^{k}$) [Fig.~\ref{fig4}(d)].

One may find different dynamics of $L$ and $D$ in Fig.~\ref{fig4}(a).
In previous our experiment~\cite{Kuroda16prl}, the decay of $D$ due to $\Gamma_{e}^{k}$ was found to be slow 2.5~ps.
The data shows that $L$ obviously decays faster than that for $D$ and the decay time of $L$ is determined to be 300~fs.
This result likely indicates different mechanisms in $\Gamma_{e}^{k}$ for these two components of the surface photocurrent.
Recently many spin-resolved photoemission experiments presented the strong polarization dependence of the spin excitation~\cite{Joswiak13NaturePhys,Zhu14prl,Xie14NatureCom,Kuroda16prb}.
Sanchez-Barriga $et\; al$ directly investigated the spin-polarization of the TSS in out-of-equilibrium by spin- and time-resolved 2PPE and found its relaxation time of nearly 300~fs~\cite{Barriga16prb}.
Thus, the transient spin polarization in the TSS, if exists in the mid-IR excitation, may enhance $\Gamma_{e}^{k}$, since the suppression of the backscattering of the spin-helical texture is softened.
The fast dynamics of $L$ can therefore include the spin-relaxation mechanism but might only be disentangled by a direct observation of the spin dynamics by using a combination of spin-resolved 2PPE and mid-IR pump pulses.
Indeed, $\Gamma_{e}^{k}$ becomes the main factor for the transport properties under static electric fields if the sample is close to charge neutrality.
Thus, the knowledge of the different dynamics of the coherent surface photocurrent in the Dirac cone is critically important for future spintornic applications.

In conclusion, we have directly investigated the ultrafast excitation and dynamics of the surface Dirac photocurrent in Sb$_{2}$Te$_{3}$ by means of time-resolved 2PPE.
The direct optical transition of the TSS induced by mid-IR excitation enables us to not only generate the TSS photocurrent but also coherently control over the surface photocurrent by the mid-IR polarization through photogalvanic effect.
Our technique directly images the temporal evolution of the surface photocurrent in the TSS in momentum-space, which visualizes the energy- and momentum-relaxation processes of transient electrons in the surface Dirac-cone.
This work finds a role in applications requiring ultrafast optical control exploiting spin-polarized surface Dirac electrons.

We gratefully acknowledge funding by the Deutsche Forschungsgemeinschaft through through projects No. HO2295/7 (SPP1666) and No. GU495/2.
K.~K. acknowledges support from JSPS Postdoctral Fellowship for Research Abroad and SFB 1083.
Partial support by the Saint Petersburg State University
(Grant Number 15.61.202.2015) is also acknowledged.

\bibliographystyle{prsty}

\end{document}